\documentclass[twocolumn,showpacs,preprintnumbers,amsmath,amssymb,footinbib,superscriptaddress]{revtex4-1}
\usepackage[utf8]{inputenc}
\usepackage[english]{babel}
\usepackage{lmodern}
\usepackage{amsmath}
\usepackage{graphicx}
\usepackage{amssymb}
\usepackage{epstopdf}
\usepackage{ upgreek }
\usepackage{color}
\usepackage{setspace}
\usepackage{bbm}
\usepackage{ mathrsfs }
\usepackage[bottom]{footmisc}

\graphicspath{{Figures/}}

\usepackage[toc,page]{appendix}
\usepackage{tocloft}

\begin{document}

\title{Dissipative stabilization of dark quantum dimers via squeezed vacuum}
\author{R.~Guti\'{e}rrez-J\'{a}uregui}
\email[Email:]{r.gutierrez.jauregui@gmail.com}
\author{A.~Asenjo-Garcia}
\affiliation{Department of Physics, Columbia University, New York, New York 10027, USA}
\author{G.~S.~Agarwal}
\affiliation{Institute for Quantum Science and Engineering, and Departments of Biological and Agricultural Engineering, and Physics and Astronomy Texas A{\&}M University, College Station, TX 77843}

\begin{abstract}

Understanding the mechanism through which an open quantum system exchanges information with an environment is central to the creation and stabilization of quantum states. This theme has been explored recently, with attention mostly focused on system control or environment engineering. Here, we bring these ideas together to describe the many-body dynamics of an extended atomic array coupled to a squeezed vacuum. We show that fluctuations can drive the array into a pure dark state decoupled from the environment. The dark state is obtained for an even number of atoms and consists of maximally entangled atomic pairs, or dimers, that mimic the behavior of the squeezed field. Each pair displays reduced fluctuations in one polarization quadrature and amplified in another. This dissipation-induced stabilization relies on an efficient transfer of correlations between pairs of photons and atoms. It uncovers the mechanism through which squeezed light causes an atomic array to self-organize and illustrates the increasing importance of spatial correlations in modern quantum technologies where many-body effects play a central role. 

\end{abstract}
\maketitle

\section{Introduction}

The collective radiation of an atomic array is an iconic example of many-body behavior in quantum open systems. It follows the loss of excitations from several atoms to a common environment, and arises from vacuum fluctuations~\cite{Dalibard_1982}. Recent interest in this process lies in the insight it provides to stabilize quantum states by protecting them from dissipation via destructive interference of individual radiative paths. The paths depend on the spatial arrangement of the array and on the spectral and spatial properties of the environment. Their manipulation builds upon a larger trend in quantum technologies: the use of spatial correlations to generate, control, and probe entangled states in extended many-body systems~\cite{Bakr_2009, Pichler_2015,Lienhard_2018,Omran_2019,Minev_2021,de_Neeve_2022,Yang_2022}.

Current interest in quantum state stabilization via correlated radiation follows two experimental trends. On the one hand, the ability to control atomic positions at the single-particle level has lead to the creation of emitter arrays whose patterns are tailored to achieve particular tasks in quantum simulation~\cite{Omran_2019,Leseluc_2019}, sensing~\cite{Madjarov_2019,Norica_2019,Ludlow_2019,Young_2020}, or information processing~\cite{Singh_2022,Kirchmair_2022}. On the other hand, fluctuations of the environment have been engineered to control the radiative response of single trapped ions and superconducting circuits~\cite{Myatt_2000,Carvalho_2001,Siddiqi_2013, Siddiqi_2016,Hoi_2015,Harrington_2019,Mendonza_2020}.

In this work we bring together ideas used to study quantum systems extended in space with those of environment engineering to describe the correlated decay of an atomic array coupled to a squeezed vacuum. Squeezed vacuum corresponds to an engineered environment composed of correlated photonic pairs~\cite{Caves_1985}. It displays a phase-sensitive amplification and deamplification of fluctuations that has been used to unveil the stochastic nature of quantum optical processes, such as spontaneous decay~\cite{Gardiner_1986,Siddiqi_2013} and resonance fluorescence~\cite{Carmichael_1987,Siddiqi_2016}. We show that---depending on the atomic positions and the spatial profile of the electromagnetic modes carrying the squeezed field---an atomic array can settle into highly entangled pure states protected from the environment. The states are built from atomic pairs that mimic the underlying environment: displaying reduced fluctuations in one polarization quadrature and amplified in another. We explore this phenomenon in one-dimensional arrays of different sizes and atomic positions to show how to manipulate the atom-atom correlations in the steady-state. Depending on the system parameters, the stabilized state is shown to be a pure dimerized state with pair-wise entanglement, a melted dimer with all-to-all correlations, or an uncorrelated mixed state. 

The paper is organized as follows. We begin in Sec.~\ref{Sec:background} by characterizing the broadband squeezed drive and deriving the atomic master equation using a cascaded-open-quantum system perspective. Then, in Sec.~\ref{Sec:collective_emission}, we map out changes in the steady state for different array separations and centers. The array is shown to decouple from the environment when atoms are placed, as pairs, at points where the two-point correlations of the field are maximized. Numerical results are supported by analytical expressions obtained via an unraveling of the master equation. The decoupled states are described in Sec.~\ref{Sec:dimerized_chain} where we introduce atom-atom interactions. The slow-fast dynamics obtained from the interplay between coherent interactions and collective dissipation are discussed in Sec.~\ref{Sec:time_scales}. Section~\ref{Sec:outlook} is left for conclusion.

\section{Background}\label{Sec:background}

The experimental realization of an artificial atom radiating into a squeezed vacuum by Siddiqi and collaborators~\cite{Siddiqi_2013,Siddiqi_2016} demonstrated the ability to tailor the environment and test the limits of conventional quantum optics using superconducting circuits~\cite{Carmichael_2016,Blais_2021}. It showed that the atom undergoes a polarization-sensitive decay, where it relaxes into a steady state following dramatically different timescales for each polarization quadrature, as predicted by Gardiner~\cite{Gardiner_1986}. Eventually, the atom reaches a mixed steady state identical to that obtained from the absorption and emission of uncorrelated thermal photons. 

Key to this observation was a source able to produce correlated photons that covered all the spatial modes surrounding the artificial atom. The atom was coupled to an environment displaying a reduced dimensionality with respect to free space to achieve this. At the time, attention was focused on the temporal correlations of the squeezed modes with their spatial structure used to determine the coupling strength and local phase of the interaction. This structure, however, plays a central role when the system includes several atoms extended in space, with each one probing a different local environment.

\subsection{Model: Atomic array and correlated travelling photons}

\begin{figure}
\includegraphics[width=.95\linewidth]{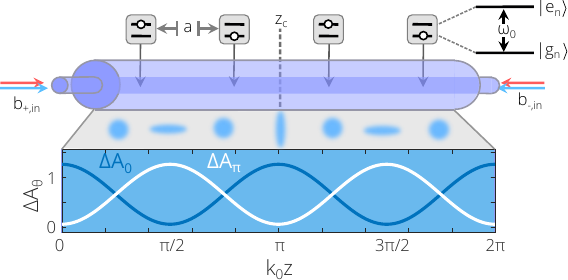}
\caption{An emitter array of lattice constant $a$ and center $z_{c}$ is coupled to a one-dimensional waveguide. The array is composed of $N_{\text{at}}$ qubits of resonance frequency $\omega_{0}$. The waveguide is driven by broadband squeezed light through two ports. Superposition of both drives causes a periodic amplification and deamplification in the field quadratures fluctuations, as shown by the local variances $\Delta A_{0}^{2}$ and $\Delta A_{\pi}^{2}$ taken from Eq.~(\ref{eq:variance}) with $\varphi=0$, $N_{\text{ph}}=0.88$, $M_{\text{ph}}^{2}=N_{\text{ph}}(N_{\text{ph}}+1)$,  and drawn as blue and white solid lines.} \label{Fig:field-variance} 
\end{figure}

We consider a one-dimensional array of $N_{\text{at}}$ atoms separated a distance $a$ from their nearest neighbors, as sketched in Fig.~\ref{Fig:field-variance}. Each atom is labelled by its position $z_{n}$ and is modelled as a two-state system with states $\vert e_{n}\rangle$ and $\vert g_{n} \rangle$ separated by a transition frequency $\omega_{0}$. The atoms couple exclusively to the electromagnetic modes of a waveguide, which acts as a source of dissipation and mediates atom-atom interactions. This waveguide has a length $L$ and is driven by a broadband squeezed source through two ports that control the input and output of photons.

The electric field inside the waveguide is naturally expanded in terms of travelling modes. Its positive frequency component reads
\begin{align}\label{eq:Fourier}
\mathcal{E}_{s}(z,t) &= \sqrt{\frac{Lc}{2\pi}}\int d\omega \int_{0}^{\infty} dk  \delta(\omega - c  k ) e^{-i\omega t + sikz} b_{s}(k) \, ,
\end{align}
where the operator $b_{s}(k)$ annihilates a photon of wavevector $k$ and frequency $\omega$ propagating along the $s=\pm$ direction. In writing Eq.~(\ref{eq:Fourier}) we have assumed a linear dispersion $\omega = c \vert k \vert$, which results in a density of modes $g(\omega) = L/2\pi c$. 

Correlations between different waveguide modes are grounded on the physical process used to produce and transmit the squeezed light~\cite{Devoret_2009,Devoret_2012}. We consider here a parametric amplifier source driving the waveguide. The source runs on a photon-photon interaction mediated by a nonlinear medium that is activated by an external pump~\cite{Caves_1987}. To be specific, we have in mind a Josephson travelling wave amplifier~\cite{Macklin_2015,Grimsmo_2017} whose nonlinearity works as an analogue to those of atomic gases~\cite{Slusher_1986, Wu_1986} or optical fibers~\cite{Shelby_1986} used in seminal experiments of squeezing. The amplifier outputs correlated pairs of photons, $a$ and $b$, whose frequencies and wavevectors follow the phase matching conditions $\omega_{a} + \omega_{b} = 2\omega_{c}$ and $k_{a} + k_{b} = 2k_{c}$ with $\omega_{c}$ the central frequency of the amplifier. The input-output relation for each pair is
\begin{equation}\label{eq:input-output}
b_{s,\text{out}}(k_{c}+k) = u_{k} b_{s,\text{in}}(k_{c}+k) + v_{k} b_{s,\text{in}}^{\dagger}(k_{c}-k)
\end{equation}
where the squeezing parameters $u_{k}$ and $v_{k}$ depend on the pump strength and nonlinear interaction. They satisfy $\vert u_{k}\vert^{2} - \vert v_{k}\vert^{2} = 1$~\cite{Grimsmo_2017}. 

The amplifier outputs photons across a broad operational bandwidth, which are fed into the waveguide. We set the central frequency of the amplifier to match the atomic resonance frequency ($\omega_{c} = \omega_{0}$) and consider a bandwidth much broader than the atomic decay rate so the squeezed drive covers all the modes relevant for the atomic interaction. The squeezed field inside the waveguide appears as $\delta$-correlated white noise characterized by 
\begin{subequations}\label{eq:correlations_2}
\begin{align}
&\langle \mathcal{E}_{s}^{\dagger}(z_{n},t_{n}) \mathcal{E}_{s}(z_{m},t_{m}) \rangle = N_{\text{ph}} \delta(\tau_{ns} - \tau_{ms})  \, , \\
&\langle \mathcal{E}_{s}(z_{n},t_{n}) \mathcal{E}_{s}^{\dagger}(z_{m},t_{m}) \rangle = (N_{\text{ph}}+1)\delta(\tau_{ns} - \tau_{ms})   \, , \\
&\langle \mathcal{E}_{s}(z_{n},t_{n}) \mathcal{E}_{s}(z_{m},t_{m}) \rangle = e^{-i\omega_{c}(\tau_{ns} + \tau_{ms})}M_{\text{ph}}\delta(\tau_{ns} - \tau_{ms})    \, , \\
&\langle \mathcal{E}_{s}^{\dagger}(z_{n},t_{n}) \mathcal{E}_{s}^{\dagger}(z_{m},t_{m}) \rangle = e^{+i\omega_{c}(\tau_{ns} + \tau_{ms})}M_{\text{ph}}^{*} \delta(\tau_{ns} - \tau_{ms})  \, , 
\end{align}
\end{subequations}
and $\langle \mathcal{E}_{s}(z,t) \rangle = 0$. Here, $c\tau_{ns} = ct_{n} - sz_{n}$ accounts for retardation and the parameters $N_{\text{ph}}$ and $M_{\text{ph}}$ follow from $u_{k}$ and $v_{k}$ of Eq.~(\ref{eq:input-output})~\cite{Gardiner_1991}. The parameters satisfy $N_{\text{ph}}(N_{\text{ph}}+1) \ge \vert M_{\text{ph}}\vert^{2}$ and reach the equality for states of minimal uncertainty~\cite{Carmichael_1987}.  

Equations~(\ref{eq:correlations_2}) describe a field with $N_{\text{ph}}$ photons per mode that are correlated in pairs through $M_{\text{ph}}$. The correlations cause a phase-dependent amplification and deamplification of fluctuations of the field quadratures
\begin{equation}
A_{\theta}(z,t) = \tfrac12 \sum_{s=\pm} (\mathcal{E}_{s}(z,t)e^{i\theta} + \mathcal{E}_{s}^{\dagger}(z,t)e^{-i \theta} ) \, .
\end{equation}
As the phase of each travelling field $\mathcal{E}_{s}(z)$ rotates while it propagates along the waveguide, the maximally squeezed quadrature of this field rotates as well. The superposition of left- and right-propagating fields then gives way to an oscillating two-point correlation  
\begin{align}\label{eq:two_point}
\langle A_{\theta}(z_{n}) A_{\theta}(z_{m}) \rangle &= \tfrac12 \vert M_{\text{ph}} \vert \cos(\theta - \varphi)\cos k_{c}(z_{n}+z_{m}) \nonumber \\
& + \tfrac12 (N_{\text{ph}}  + \tfrac12) \, ,
\end{align}
where the reference phase $\varphi$ is set by $M_{\text{ph}} = \vert M_{\text{ph}} \vert e^{i \varphi}$. The local variances in field quadratures 
\begin{equation}\label{eq:variance}
\Delta A_{\theta}^{2}(z) = \langle \delta A_{\theta}(z) \delta A_{\theta}(z) \rangle \, ,
\end{equation} with $\delta A_{\theta} = A_{\theta} - \langle A_{\theta} \rangle$, reflect an spatial dependence of fluctuations in this background field, as shown in Fig.~\ref{Fig:field-variance} for maximally and minimally squeezed quadratures.

\subsection{Master equation for an atomic array submerged in an squeezed environment}

The waveguide modes carry the squeezed drive and photons scattered in and out of the array. This composite field is, in general, non-classical and is described by correlation functions of many orders. We follow the theory of cascaded quantum systems~\cite{Gardiner_1993, Carmichael_1993} to model the evolution of the field sources and derive the master equation for an array radiating into a squeezed vacuum.

In the electric-dipole and rotating-wave approximations each atom probes the amplitude of the local electromagnetic field via the interaction Hamiltonian
\begin{equation}\label{eq:sys_reservoir_interaction}
\mathcal{H}_{SR} = \hbar \sum_{n,s} \sqrt{\gamma_{s}} (\mathcal{E}_{s}(z_{n})\sigma_{+}^{(n)} + \mathcal{E}^{\dagger}_{s}(z_{n}) \sigma_{-}^{(n)} ) \, ,
\end{equation}
where $\sigma^{(n)}_{+} = \vert e_{n} \rangle \langle g_{n} \vert$ and $\sigma^{(n)}_{-} = \vert g_{n} \rangle \langle e_{n} \vert$ are raising and lowering operators for the $n$th-atom; and $\gamma_{s}=\tfrac12\gamma$ with $\gamma$ the decay rate into the waveguide~\cite{us_Markov}. The total field is composed of free and scattered components
\begin{equation}\label{eq:field_total}
\mathcal{E}_{s}(z,t) = \mathcal{E}_{\text{f}s}(z,t) -i \sum_{n,s} \sqrt{\gamma_{s}} \sigma_{-}^{(n)}(t-st^{\prime}_{n}) \Theta(t-st^{\prime}_{n}) \, ,
\end{equation}
obtained from the Heisenberg equation of motion using Eqs.~(\ref{eq:Fourier}) and~(\ref{eq:sys_reservoir_interaction}). Here $ct_{n}^{\prime}=(z-z_{n})$ describes a time delay between emission and absorption of an excitation and the step function $\Theta(x)$ ensures causality. 

Equation~(\ref{eq:field_total}) accounts for the evolution of a source as its output field reaches its target. The spatial separation between source and target is effectively removed by moving into an interaction picture where the sources are retarded. For a small array---such that the only changes on the source as the field propagates from one end of the array to the other are given from its free evolution---this retardation produces a local phase only~\cite{Lehmberg_1970,Carmichael_1993}. The Schr\"{o}dinger picture operators ($t=0$) of the field under this assumption are
\begin{align}\label{eq:field_total_Markov}
&\mathcal{E}_{+}(z_{n}) = e^{ik_{0} (z_{n} - z_{1})}\mathcal{E}_{\text{f}+}(z_{1}) -i \tfrac12 \sqrt{\gamma_{+}}  \sum_{m<n} e^{ik_{0}(z_{n}-z_{m})}\sigma_{-}^{(m)} \, , \nonumber \\
&\mathcal{E}_{-}(z_{n}) = e^{ik_{0} (z_{N_{\text{at}}}-z_{n})}\mathcal{E}_{\text{f}-}(z_{N_{\text{at}}}) -i \tfrac12 \sqrt{\gamma_{-}}  \sum_{m>n} e^{ik_{0}(z_{m}-z_{n})}\sigma_{-}^{(m)} \, ,
\end{align}
where we have divided into right- and left-propagating channels to define source and target consistently.

The master equation for the density matrix of the array $\rho$ is derived by substituting Eq.~(\ref{eq:field_total_Markov}) into Eq.~(\ref{eq:sys_reservoir_interaction}) and following the standard approach~\cite{Agarwal_1974,Carmichael_2008}. The amplitude and correlations of the free, squeezed field are traced back to its value at the edges of the array. In an interaction picture with respect to the free term $\sum_{n} \hbar \omega_{0} \sigma^{(n)}_{+}\sigma^{(n)}_{-}$, the master equation reads
\begin{widetext} 
\begin{align}\label{eq:master_2}
\dot{\rho} = \frac{1}{i\hbar}\left[ \mathcal{H}_{\text{scatt}}, \rho \right] &+ \tfrac12\gamma \left( (N_{\text{ph}}+1) \mathcal{L}_{J_{+}}\rho +  N_{\text{ph}} \mathcal{L}_{J_{+}^{\dagger}}\rho  + \tfrac12 \vert M_{\text{ph}}\vert  \mathcal{L}_{J_{\varphi,+}}\rho - \tfrac12 \vert M_{\text{ph}}\vert  \mathcal{L}_{J_{\varphi+\pi,+}}\rho  \right) \, \nonumber \\
&+ \tfrac12\gamma \left( (N_{\text{ph}}+1) \mathcal{L}_{J_{-}}\rho +  N_{\text{ph}} \mathcal{L}_{J_{-}^{\dagger}}\rho  + \tfrac12 \vert M_{\text{ph}}\vert  \mathcal{L}_{J_{\varphi,-}}\rho - \tfrac12 \vert M_{\text{ph}}\vert  \mathcal{L}_{J_{\varphi+\pi,-}}\rho  \right) \, ,
\end{align}
\end{widetext} 
where atom-atom interactions via the two counter-propagating channels $s=\pm$ sum to give 
\begin{equation}\label{eq:Hamiltonian}
\mathcal{H}_{\text{scatt}} = \tfrac12 \hbar\gamma \sum_{n,m=1}^{N_{\text{at}}} \sin  k_{0}\vert z_{n}-z_{m} \vert \sigma_{+}^{(n)}\sigma_{-}^{(m)} \, .
\end{equation}
Loss is accounted for by $\mathcal{L}_{\xi} \cdot \equiv \xi\cdot\xi^{\dagger} - \cdot\tfrac12\xi^{\dagger}\xi - \tfrac12\xi^{\dagger}\xi\cdot$ with collective jump operators 
\begin{subequations}\label{eq:jumps_1}
\begin{align}
&J_{s} = \sum_{n}e^{-isk_{0}z_{n}} \sigma^{(n)}_{-} \, , \\
&J_{\varphi,s} = e^{i\varphi/2}J_{s}+e^{-i\varphi/2} J_{s}^{\dagger} \, ,
\end{align}
\end{subequations}
whose reference phase $\varphi$ follows the from squeezed-light source, defined below Eq.~(\ref{eq:two_point}). We take  $\varphi = 0$ throughout.

\section{Correlated decay channels and stabilization}\label{Sec:collective_emission}

Equation~(\ref{eq:master_2}) sets out a model for the collective emission of an atomic array coupled to a squeezed vacuum. It is written in a form that highlights collective effects over two-photon processes of Refs.~\cite{Zubairy_2018} and~\cite{Bai_2021}. Like the single atom case, the engineered fluctuations translate into phase-sensitive radiative decays through the jump operators $J_{\varphi, s}$ and $J_{\varphi+\pi, s}$. But, with several atoms spread along the waveguide and a spatially changing background, each atom may probe different squeezed quadratures. 

\begin{figure}
\begin{center}
\includegraphics[width=.95\linewidth]{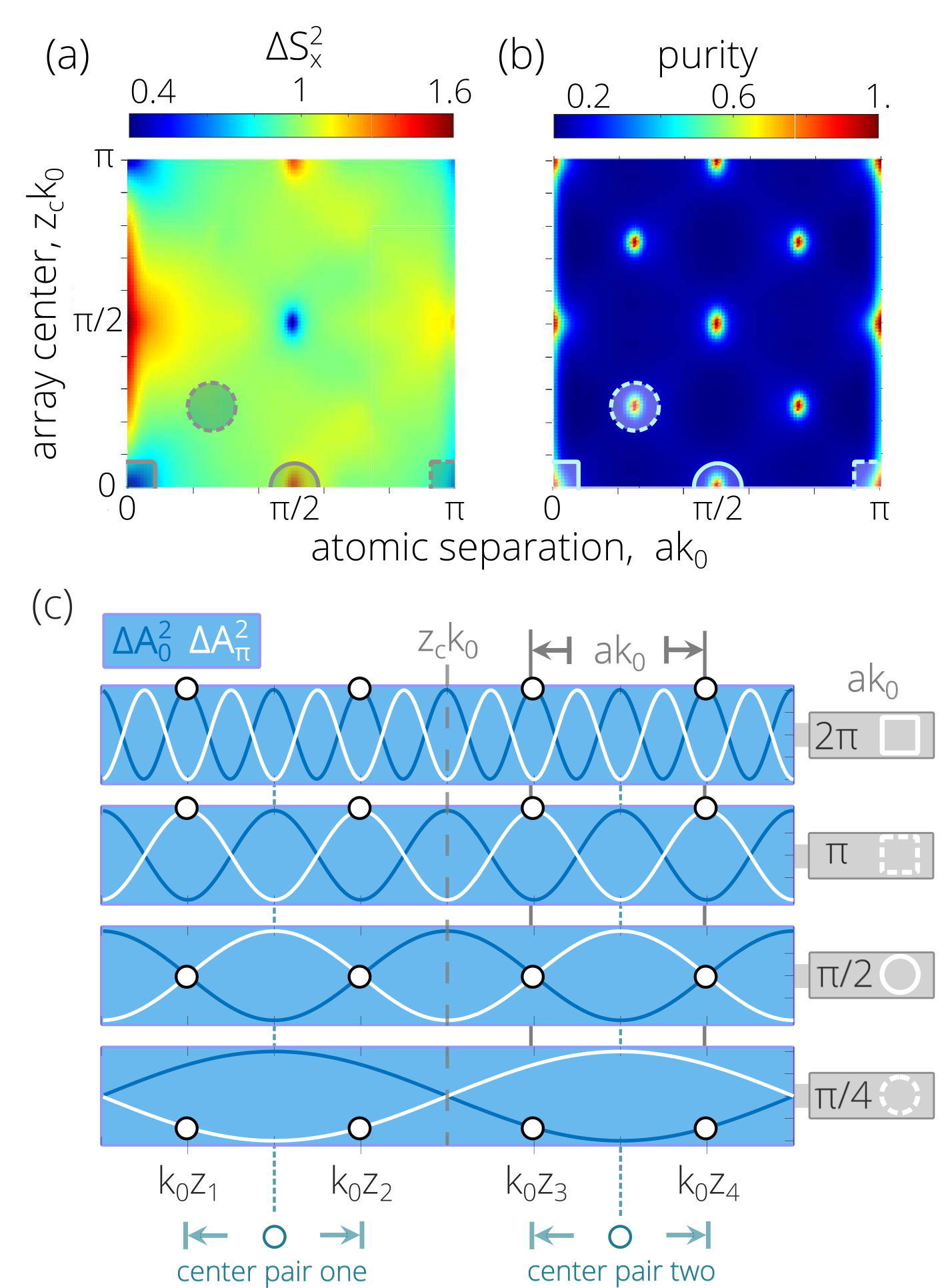}
\caption{Steady state variance $\Delta S_{x}^{2}$ and purity of an atomic array as a function of its center and atomic separation. The array has $N_{\text{at}}=4$ atoms coupled to a perfectly squeezed vacuum with $N_{\text{ph}}=0.88$ and $\varphi=0$. (a) Fluctuations in the polarization quadrature $S_{x}$ are amplified and deamplified at points where all atoms can be arranged into pairs that satisfy $\cos k_{0}(z_{n}+z_{m})=\pm 1$, a condition that follows from the two-field correlation of the field quadratures in Eq.~(\ref{eq:two_point}). (b) The system is pure at these points and mixed away from them. (c) Atomic positions (circles) relative to the local variance in field quadratures $\Delta A_{0,\pi}$ (white and blue lines) for cases where the array reaches a pure state. These examples represent the points signalled by squares and circles in (a)-(b). In all cases sequential atomic pairs are centered at points where the local variance in field quadratures maximizes (dark blue circles). } \label{Fig:variance_ss} 
\end{center}
\end{figure}

In Figure~\ref{Fig:variance_ss} we map out changes in the steady state $\rho_{ss}$ as a function of array center $z_{c}$ and lattice constant $a$. The array has $N_{\text{at}}=4$ atoms and is submerged in a perfectly squeezed vacuum [$\vert M_{\text{ph}} \vert^{2} = N_{\text{ph}}(N_{\text{ph}}+1)$] with an average of $N_{\text{ph}}=0.88$ photons per mode. All atoms begin in the ground state and are set to evolve numerically using Eq.~(\ref{eq:master_2}) until a steady state is reached. The collective state is tracked through the polarizations
\begin{align}\label{polarization_quadratures}
S_{j} = \tfrac12 \sum_{n=1}^{N_{\text{at}}} \sigma^{(n)}_{j}  \, ,
\end{align}
with $j= \lbrace x,y,z\rbrace$ and $\sigma_{\pm}^{(n)} = \sigma^{(n)}_{x} \pm i \sigma^{(n)}_{y}$, and through its purity $\text{Tr}[\rho_{ss}^{2}]$. We plot the variance $\Delta S_{x}^{2}$ in Fig.~\ref{Fig:variance_ss}(a) to explore how field correlations transfer into the atoms. This variance changes as we move along the parameter space and reaches a local extrema at particular points as those signaled in the figure. The steady state is pure at these points $(\text{Tr}[\rho_{ss}^{2}] =1)$, suggesting that the system is trapped in a dark state decoupled from the environment. These points correspond to particular atomic arrangements that we divide into four cases drawn in Fig.~\ref{Fig:variance_ss}(c). The cases share the commonality that all atoms can be organized into pairs that satisfy $\cos k_{0}(z_{n}+z_{m}) = \pm 1$. Atomic pairs placed at these positions probe a maximally correlated field quadrature, as seen from the two-point correlation of Eq.~(\ref{eq:two_point}). By plotting the atomic positions relative to the background field in Fig.~\ref{Fig:variance_ss}(c), we show that this condition places the center of each atomic pair at points where the local variance of field quadratures is maximally squeezed.

\subsection{Pointer states}

The physical origin of the dark states follows the exchange of information between field and atoms. This exchange depends on the particular set of jump operators $J_{i}$ used to unravel the master equation. The pointer states $\vert \psi_{i} \rangle$ of a jump operator $J_{i}$ are defined as~\cite{Brune_1996,Carvalho_2001,Zurek_1981,Haroche_2006}
\begin{equation}\label{eq:definition_pointer}
J_{i} \vert \psi_{i} \rangle = \lambda_{i}\vert \psi_{i} \rangle \, .
\end{equation}
Pointer states do not entangle with the environment under the action of $J_{i}$. We can then define a dark state as a pointer state with eigenvalue $\lambda_{0}=0$ for all jump operators of a given unravelling. These dark states evolve without any influence from the environment and only experience the Hamiltonian $\mathcal{H}_{\text{scatt}}$ of Eq.~(\ref{eq:Hamiltonian})~\cite{Haroche_2006}. 

In this Section we focus on the collective jump operators and possible unravelings of the master equation to identify the dark states. We return to the effect of the Hamiltonian in Sec.~\ref{Sec:dimerized_chain} where we present its role to stabilize the collective state of the array. 

\subsection{Unravelling of the master equation}

A natural unraveling of Eq.~(\ref{eq:master_2}) tracks the exchange of excitations between array and environment through the jump operators $J_{s}$ and $J_{\phi,s}$ of Eq.~(\ref{eq:jumps_1}). When seen as independent processes, each jump operator represents a different way for the environment to acquire information from the system. Operators $J_{s}$ and $J_{s}^{\dagger}$ track photon absorption and emission as the array exchanges excitations with left- and right-propagating channels; while operators $J_{\varphi,s}$ and $J_{\varphi+\pi,s}$ track changes in phase through two photon exchange.  

Photons in a squeezed environment, however, are created as correlated pairs and a particular unravelling of the master equation can be constructed to highlight this feature. We begin to construct this unravelling by defining the standing-wave operators
\begin{subequations}\label{eq:collective_decay_channels}
\begin{align}
{S}_{\pm}^{(R)} = \sum_{n=1}^{N_{\text{at}}} \cos k_{0}z_{n} {\sigma}_{\pm}^{(n)} \, , \\
{S}_{\pm}^{(I)} = \sum_{n=1}^{N_{\text{at}}} \sin k_{0}z_{n} {\sigma}_{\pm}^{(n)} \, ,
\end{align}
\end{subequations}
connected to the travelling jump operators through
\begin{equation}\label{eq:standing_to_traveling}
J_{s} = S^{(R)}_{-} - is S^{(I)}_{-}  \, .
\end{equation}
The standing wave sets two decay channels separated by a phase of $\pi$. Each channel aligns with a phase quadrature of the field. Then, in the limit of perfect squeezing, the jump operators can be further arranged into
\begin{subequations}\label{eq:squeezed_operators}
\begin{align}
{\mathcal{J}}_{x} = \frac{\mu S_{-}^{(I)} + \nu S_{+}^{(I)} }{\vert 4 \mu \nu \vert^{1/2}} \, , \\
{\mathcal{J}}_{y} = \frac{\mu S_{-}^{(R)} - \nu S_{+}^{(R)} }{\vert 4 \mu \nu \vert^{1/2}}\, ,
\end{align}
\end{subequations}
where $\mu$ and $\nu$ relate to the squeezing parameters by
\begin{align}\label{eq:conditions_operators}
&\vert \mu \vert^{2} = (N_{\text{ph}}+1) \, , \vert \nu \vert^{2} = N_{\text{ph}} \, , \nu \mu^{*} = - M_{\text{ph}}  \, ,
\end{align}
and satisfy $\vert \mu \vert^{2} - \vert \nu \vert^{2} = 1$.

Using the linear transformations of Eqs.~(\ref{eq:collective_decay_channels})-(\ref{eq:squeezed_operators}), the master equation can be written using two jump operators only
\begin{align}
\dot{\rho} = -i\hbar^{-1}\left[ \mathcal{H}_{\text{scatt}}, \rho \right] + 4 \gamma \vert \mu \nu \vert ( \mathcal{L}_{\mathcal{J}_{x}} \rho + \mathcal{L}_{\mathcal{J}_{y}} \rho )  \, .
\end{align}
The condition to find a dark state simplifies into
\begin{equation}\label{eq:condicion}
\det [\mathcal{J}_{x}^{\dagger}\mathcal{J}_{x} + \mathcal{J}_{y}^{\dagger}\mathcal{J}_{y}] = 0 \, ,
\end{equation}
as obtained from Eq.~(\ref{eq:definition_pointer}) with $\lambda_{x} = \lambda_{y} = 0$. 

Condition~(\ref{eq:condicion}) cannot be fulfilled by a single two-state atom. For a single atom the polarization changes between $\pm 1$ values after each jump~\cite{us_2022_b2}. When averaged out, these sudden changes lead to a mixed state density matrix. To find states that decouple from the environment we need to go beyond the single atom case. After all, with two photon processes being central to generate squeezed light, it is expected for atomic pairs to play a significant role.

\subsection{Building blocks for dark states: An atomic pair}

Here, we lay down the connection between dark states and the squeezed environment. The connection is based on an array of just two atoms and extended to an arbitrary number of atoms in Sec.~\ref{Sec:dimerized_chain}. 

Solving Eq.~(\ref{eq:condicion}) for $N_{\text{at}}=2$ shows two scenarios where the array decouples from the environment. In the first scenario atoms are separated a distance $a$ such that $\sin k_{0}a = 0$. The pair decouples from the environment by settling into the maximally entangled state
\begin{equation}\label{eq:dimer_sym}
\vert \psi^{(0)}_{nm} \rangle = \frac{\vert g_{n} e_{m} \rangle -e^{ik_{0}a} \vert e_{n} g_{m} \rangle}{\sqrt{2}} \, .
\end{equation}
Atoms separated by this distance probe the same local environment. The jump operators $\mathcal{J}_{x,y}$ reflect this fact and become symmetric ($k_{0}a=2m \pi$) or antisymmetric ($k_{0}a=(2m+1) \pi$) under atomic exchange, thus organizing the dynamics into two separate subspaces. In accordance with the master equation results of Ref.~\cite{Zubairy_2018}.

In the second scenario atoms are placed at positions $z_{n}$ and $z_{m}$ where correlations of the field quadratures maximize, \textit{i.e.}, $\cos k_{0}(z_{n}+z_{m}) = \pm 1$. At these points the dark state is
\begin{equation}\label{eq:dimer_ri1}
\vert \psi^{(1)}_{nm} \rangle = \frac{\mu \vert g_{n}g_{m} \rangle + e^{ik_{0}(z_{n}+z_{m})}\nu \vert e_{n}e_{m} \rangle}{\sqrt{\vert \mu \vert^{2} + \vert \nu \vert^{2}}}  \, .
\end{equation} 
It displays an imbalance between ground and excited states that follows from the jump operators of Eq.~(\ref{eq:squeezed_operators}). Atoms in this dark state mimic the squeezed vacuum, displaying amplified and deamplified fluctuations 
\begin{subequations}\label{eq:var_dimers}
\begin{align}
\Delta S_{x}^{2} &= \frac{\vert \mu + e^{ik_{0}(z_{n}+z_{m})}\nu \vert^{2}}{2 (\vert \mu\vert^{2} + \vert \nu\vert^{2}) } \, , \\
\Delta S_{y}^{2} &= \frac{\vert \mu - e^{ik_{0}(z_{n}+z_{m})}\nu\vert^{2}}{{2 (\vert \mu\vert^{2} + \vert \nu\vert^{2}) }} \, , 
\end{align}
\end{subequations}
and a population imbalance 
\begin{equation}\label{eq:exp_dimers}
\langle S_{z} \rangle = \frac{\vert \nu\vert ^{2} - \vert \mu\vert^{2}}{\vert \mu\vert^{2} + \vert \nu\vert^{2}} = \frac{-1}{2N_{\text{ph}}+1} \, .
\end{equation}
The state is one of minimal uncertainty on total angular momentum $\Delta S^{2}_{x} \Delta S^{2}_{y} = \langle \tfrac12 S_{z} \rangle^{2} $.  

\section{A dimerized chain}\label{Sec:dimerized_chain}

Two atoms decouple from the environment by pairing into maximally entangled states when placed at points where the two-point correlations of the field quadratures maximize. Due to the linearity of the jump operators, an array decouples from the environment when all atoms are organized into pairs that satisfy this condition. For an even number of atoms, a dimerized state constructed from the pairs of Eq.~(\ref{eq:dimer_ri1}) takes the form
\begin{equation}\label{eq:dimer_ri}
\vert \psi_{\ell} \rangle =  \bigotimes \vert \psi^{(1)}_{n_{\ell}m_{\ell}} \rangle \, ,
\end{equation}
where the product extends over all pairs. The $\ell$ index characterizes the different ways atoms can pair.

So far, we have constructed the dark state using the jump operators only and neglected the effect of the Hamiltonian $\mathcal{H}_{\text{scatt}}$ written in Eq.~(\ref{eq:Hamiltonian}). This interaction can drive the atoms out of the dark states and lead to a loss of coherence. We now provide the conditions to find stable dark states that decouple entirely from the environment and show two different behaviors that arise from this interaction. 

\begin{figure}
\begin{center}
\includegraphics[width=1.\linewidth]{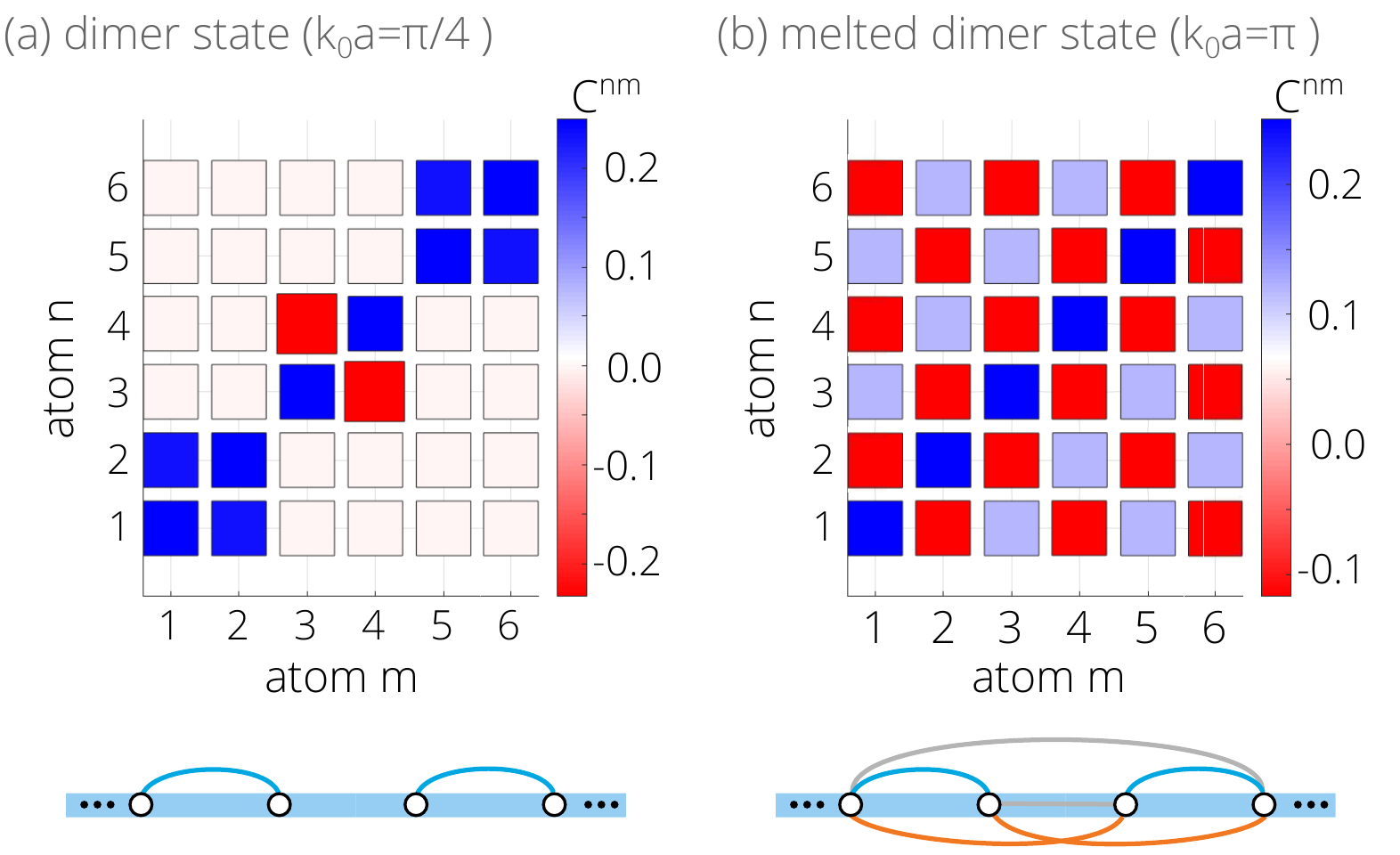}
\caption{Atomic correlations $C^{nm} = \tfrac14 \langle \sigma_{x}^{(n)} \sigma_{x}^{(m)} \rangle$ in the steady state for two array patterns. (a) For $k_{0}a=\pi/4$ the interaction $\mathcal{H}_{\text{scatt}}$ causes the atoms to dimerize following Eq.~(\ref{eq:dimer_tot}). (b) For $k_{0}a = \pi $ there is no interaction $\mathcal{H}_{\text{scatt}}$ and the steady state is a superposition of all possible dimers, thus displaying long-range entanglement between all atoms. In both cases the array has $N_{\text{at}}=6$ atoms initially prepared in the ground state, it is centered at $k_{0}z_{c} = 0$, and is coupled to a perfectly squeezed vacuum with $N_{\text{ph}}=0.88$, $\varphi=0$.} \label{Fig:correlations} 
\end{center}
\end{figure}

\subsection{Dimerized chain ($\sin ak_{0} \neq 0$)}\label{Sec:dimerized_chain_one}

Due to the long-range interactions inside the waveguide, atoms couple via $\mathcal{H}_{\text{scatt}}$ for distances $ak_{0} \neq m \pi$. This brings two particular pairs, $\vert \psi^{(1)}_{a_{1}a_{2}} \rangle$ and $\vert \psi^{(1)}_{b_{1}b_{2}} \rangle$, out of the dimerized state of Eq.~(\ref{eq:dimer_ri}) and into the same final state $\vert e_{a1} g_{a2} e_{b1} g_{b2} \rangle$. The probability amplitude to leave the dimer following this process is
\begin{align}
c_{\text{esc}} \propto e^{i2k_{0}\bar{z}_{a}} \sin k_{0} \vert z_{a2} - z_{b1} \vert +e^{i2k_{0}\bar{z}_{b}} \sin k_{0} \vert z_{a1} - z_{b2} \vert \nonumber
\end{align}
where $\bar{z}_{\ell}= (z_{\ell1}+z_{\ell2})/2$ is the center of the $\ell$-pair. 

The array remains trapped in a dimer state if all possible paths out of it interfere destructively. It can be shown that such destructive interference occurs when the atoms only pair with their nearest neighbor, starting from the boundaries~\cite{us_2023_App}. The only possible product of dimers immune to the Hamiltonian interaction is thus
\begin{equation}\label{eq:dimer_tot}
\vert \psi \rangle = \bigotimes_{n = 1}^{N_{\text{at}}/2} \frac{\mu \vert g_{2n-1}g_{2n} \rangle + e^{ik_{0}(z_{2n-1}+z_{2n})}\nu \vert e_{2n-1}e_{2n} \rangle}{\sqrt{\vert \mu \vert^{2} + \vert \nu \vert^{2}}}  \, , 
\end{equation}
where $\cos k_{0}(z_{2n-1}+z_{2n}) = \pm 1$. Since the atoms are equidistant, stable dark states exist for separations $k_{0}a=m\pi/4$ only, with the coherent interaction acting for $m\mod 4 \ne 0$. This separation is in accordance with Fig.~\ref{Fig:variance_ss}. 

Atoms in this dark state are maximally entangled with their partner and uncorrelated from the rest. This is shown in Fig.~\ref{Fig:correlations}(a) where we plot the atomic correlation 
\begin{equation}\label{eq:corr}
C^{nm} = \tfrac14 \langle \sigma_{x}^{(n)} \sigma_{x}^{(m)} \rangle \, 
\end{equation}
for an array of six atoms separated a distance $k_{0}a=\pi/4$. The correlations are obtained by evolving the system numerically from its ground state using Eq.~(\ref{eq:master_2}), and show excellent agreement with the analytical result. Correlations change sign from pair to pair as successive centers are placed at points where the variance $\Delta A_{0}$ goes from squeezed to antisqueezed. This flipping reduces the fluctuations of the total polarization 
\begin{subequations}\label{eq:var_dimers_tot}
\begin{align}
\Delta S_{x}^{2} &= \frac{\sum_{n}\vert \mu + e^{ik_{0}(z_{2n-1}+z_{2n})}\nu\vert^{2}}{N_{\text{at}} (\vert \mu\vert^{2} + \vert \nu\vert^{2}) } \, , \\
\Delta S_{y}^{2} &= \frac{\sum_{n}\vert\mu - e^{ik_{0}(z_{2n-1}+z_{2n})}\nu\vert^{2}}{N_{\text{at}} (\vert \mu\vert^{2} + \vert \nu\vert^{2}) } \, , 
\end{align}
\end{subequations}
which is obtained by adding the variances of each pair given in Eq.~(\ref{eq:var_dimers}). The same steady state is reached for every initial state.

\subsection{Melting of the dimer ($\sin k_{0}a = 0$)}

For atoms separated by a distance $k_{0}a = m\pi$ there is no coherent interaction $\mathcal{H}_{\text{scatt}}$ to drive the system out of the dark states or to distinguish between different possible dimers. All atoms probe the same background field and can pair with one another. 

In Figure~\ref{Fig:correlations}(b) we plot the correlations in steady state for an array of six atoms separated a distance $k_{0}a = \pi$. The steady state is highly entangled and displays long-range correlations that change in sign from pair to pair. The sign follows from a sum over all possible pairings using Eqs.~(\ref{eq:dimer_ri1}) and~(\ref{eq:dimer_ri}). This state is constructed from the two-atom dark states $\vert \psi^{(1)}_{nm} \rangle$ [Eq.~(\ref{eq:dimer_ri1})], but, for $k_{0}a = m\pi$, additional dark states $\vert \psi^{(0)}_{nm} \rangle$ [Eqs.~(\ref{eq:dimer_sym})] emerge. The additional states found at these separations cause the array to organize within $(\tfrac12 N_{\text{at}} +1)$ separate subspaces. Each subspace is determined by the number of pairs in $\vert \psi^{(1)}_{nm} \rangle$ against those in $\vert \psi^{(0)}_{nm} \rangle$. For example, the dark states of four atoms are
\begin{align}
&\vert \psi_{\text{dark},2} \rangle \propto \vert \psi^{(1)}_{12} \rangle\vert \psi^{(1)}_{34} \rangle +  \vert \psi^{(1)}_{13} \rangle\vert \psi^{(1)}_{24} \rangle +  \vert \psi^{(1)}_{14} \rangle\vert \psi^{(1)}_{23} \rangle \, , \nonumber \\
&\vert \psi_{\text{dark},1} \rangle \propto \vert \psi^{(1)}_{12} \rangle\vert \psi^{(0)}_{34} \rangle +  \vert \psi^{(1)}_{13} \rangle\vert \psi^{(0)}_{24} \rangle +  \vert \psi^{(1)}_{14} \rangle\vert \psi^{(0)}_{23} \rangle + 0 \leftrightarrow 1  \, , \nonumber \\
&\vert \psi_{\text{dark},0} \rangle \propto \vert \psi^{(0)}_{12} \rangle\vert \psi^{(0)}_{34} \rangle +  \vert \psi^{(0)}_{13} \rangle\vert \psi^{(0)}_{24} \rangle +  \vert \psi^{(0)}_{14} \rangle\vert \psi^{(0)}_{23} \rangle \, . \nonumber 
\end{align}
This organization can be shown to hold for small arrays $(N_{\text{at}} = 2,4,6)$ and is expected to extend for larger arrays given the linearity of the jump operators. Since Fig.~\ref{Fig:correlations}(b) is obtained by evolving an array from its ground state, the initial state overlap with $\vert \psi_{\text{dark}, N_{\text{at}}/2} \rangle$ only.

Using these subspaces, it is possible to expand the dark states within a collective basis $\vert l, n_{e} \rangle$ with $l$ the number of pairs and $n_{e}$ the excitations. The states take the form 
\begin{equation}\label{eq:angular_mom}
\vert \psi_{\text{dark},l} \rangle = \sum_{n_{e}=0}^{N_{\text{at}}} e^{-\eta (n_{e}-N_{\text{at}}/2)} c_{l,n_{e}} \vert l, n_{e} \rangle \, ,
\end{equation}
where each excitation is weighted by a probability amplitude $c_{l,n_{e}}$ and the Lorentz parameter $e^{-\eta}  \vert \mu \vert = e^{\eta}  \vert \nu \vert$~\cite{us_2022_b2}. The amplitudes account for a statistical weight obtained from the possible dimerizations. For $l=\tfrac12 N_{\text{at}}$ pairs, the amplitudes for even excitations are
\begin{equation}\label{eq:prob_amplitudes}
c_{l,n_{e}} =  \frac{ \left( n_{g}! n_{e}! \right)^{1/2}}{2^{N_{at}/2}(\tfrac12 n_{g})!(\tfrac12 n_{e})!} 
\end{equation}
and zero otherwise~\cite{us_2022_f}. Here, $n_{g} = N_{\text{at}}-n_{e}$.

The dark states $\vert \psi_{\text{dark},l} \rangle $ were obtained by Agarwal and Puri~\cite{Agarwal_1991} for $k_{0}a = 2m\pi$ and $k_{0}z_{c}=\{ 0,\pi/2 \}$, and have reemerged recently for their potential applications on metrology~\cite{Bai_2021,Clerk_2022}. Their connection to the underlying dimers has been, however, unexplored. At these points one of the jump operators $\mathcal{J}_{x,y}$ cancels while the other can be transformed to the polarization $S_{x,y}$ via a Lorentz-like transformation~\cite{us_2022_b2}. The dark states are then obtained by rotating the eigenstates of zero angular momentum projection, $\vert l, m=0 \rangle $, and inverting the Lorentz transformation. They take the explicit form~\cite{us_2022_d}
\begin{equation}\label{eq:collective_agarwal}
\vert \psi_{\text{dark},l} \rangle = \sqrt{\frac{4 \pi}{2l +1}} \sum_{m=-l}^{l} e^{-\eta m} (-1)^{(l+m)/2} Y_{l,m}(\tfrac{\pi}{2},0) \vert l, m \rangle \, , \
\end{equation}
with $Y_{l,m}$ the spherical harmonics. Equation~(\ref{eq:collective_agarwal}) can be shown to be a special case of Eq.~(\ref{eq:angular_mom}) using identities of the spherical harmonics~\cite{Abramowitz_1972}. 

The selection of a particular dimer out of the melted state via $\mathcal{H}_{\text{scatt}}$ and $\mathcal{J}_{x,y}$ is reminiscent of a protocol used in chiral waveguides to stabilize entangled states~\cite{Pichler_2015}. The protocol considers arrays of separation $ak_{0}=2\pi n$, radiating into regular vacuum and driven by coherent fields. It relies on the existence of only one decay channel whose dark states span a large subspace. Then, by tuning the driving fields the array is brought to a particular state of this dark subspace. By comparison, an array radiating into a squeezed vacuum is brought to a dark state via the underlying fluctuations. Dimers are a product of two-photon correlations probed by the atomic positions rather than imparted by an external coherent interaction. 

\subsection{Odd number of atoms}\label{Sec:discussion}

For arrays with an odd number of atoms the correlations are still transferred in pairs, leaving an extra atom in an indeterminate state. This frustration causes the array to settle into a mixed state, as can be shown by plotting the steady state purity in analogy to Fig.~\ref{Fig:variance_ss}. While not shown here, the purity reaches a maxima when the atoms are separated a distance of $\pi/4$ and centered around a maxima of the field fluctuations. The steady state can also be shown to be formed predominantly of atomic pairs plus an uncoupled atom.  

\section{Timescales and polarization sensitive decay }\label{Sec:time_scales}

\begin{figure}
\includegraphics[width=1.\linewidth]{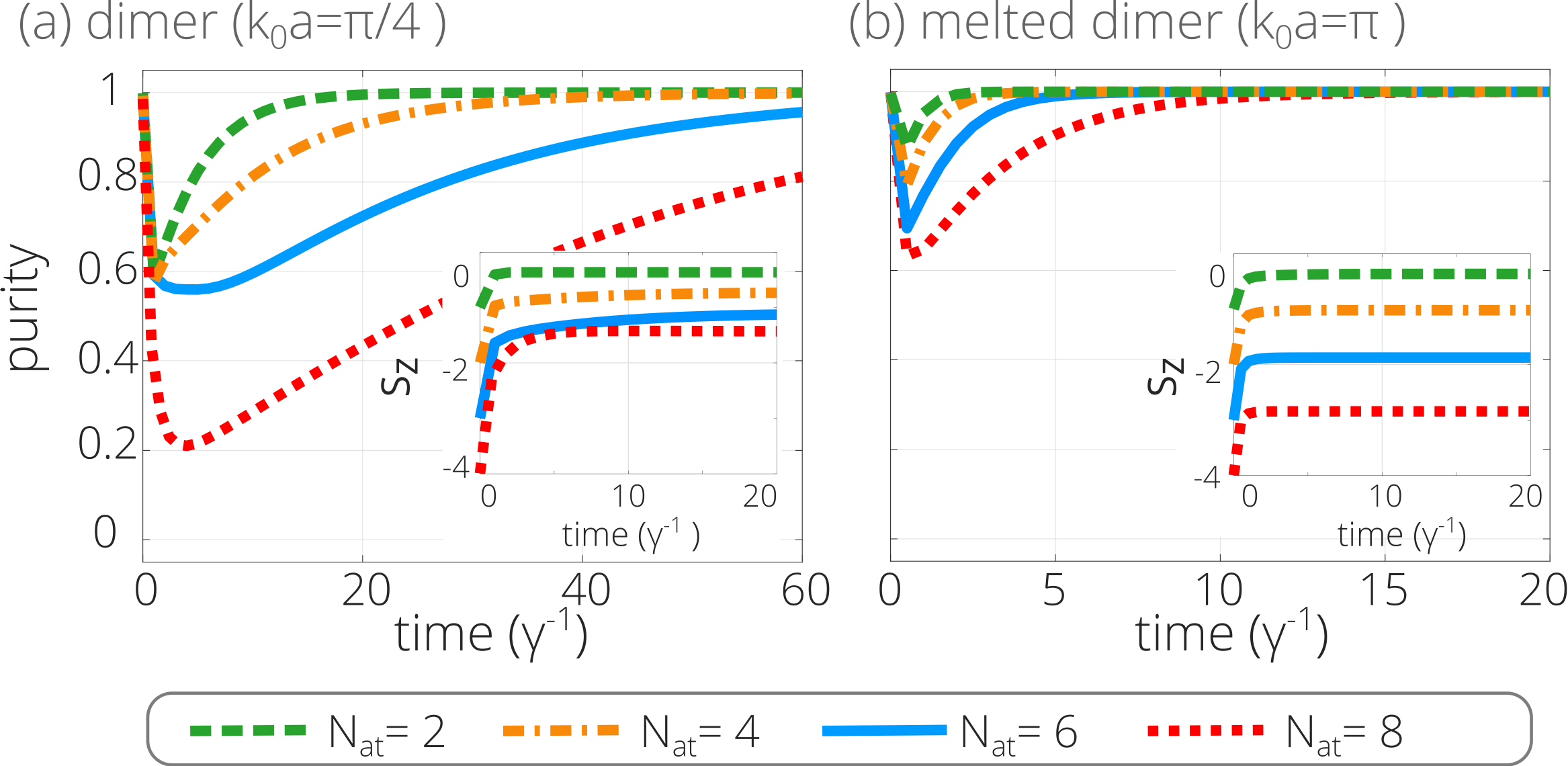}
\caption{Time evolution for different array sizes. An array prepared in the ground state settles in a steady state following two different timescales for the purity and population $\langle S_{z} \rangle$ (inset). (a) For a dimer state, the action of the Hamiltonian and two non-commuting jump operators cause the system to self-organize slowly. (b) For a melted dimer, the array quickly settles in the dark state. In all cases $N_{\text{ph}}=0.88$ and $\varphi=0$. } \label{Fig:time_scales} 
\end{figure}

\begin{figure*}
\includegraphics[width=.9\linewidth]{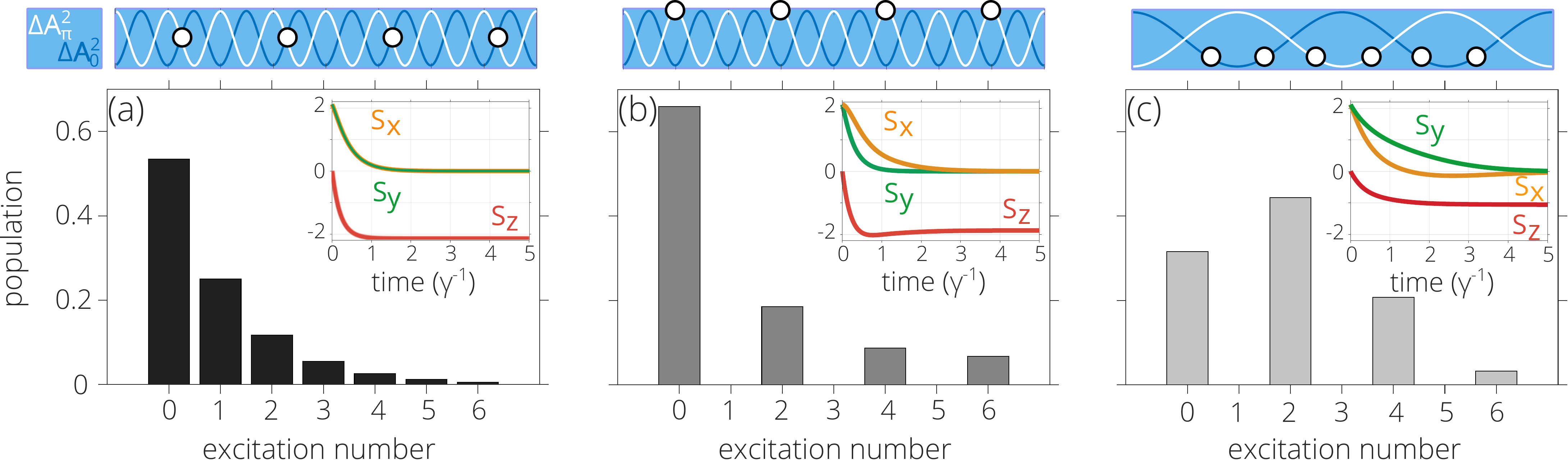}
\caption{Steady-state population and collective dynamics (inset) for three arrays with different atomic positions. Above each graph, atomic positions are sketched as circles with the local field variance $\Delta A^{2}_{0,\pi}(z)$ drawn as blue and white solid lines as a reference. (a) For center $k_{0}z_{c}=\pi/4$ and separation $k_{0}a=2\pi$, atoms probe uncorrelated photons and the polarization quadratures $S_{x}$ (yellow line) and $S_{y}$ (green) decay at the same rate. The array settles into a thermal state where all numbers of excitations are populated. (b) As the center moves to $k_{0}z_{c}=0$ all atoms are placed, as pairs, at positions where field quadratures are maximally correlated. The expected value of the polarization $\langle S_{x} \rangle$ decays at an enhanced rate while that of $\langle S_{y} \rangle$ at an inhibited one. The array settles into a correlated state where only even number of excitations are populated.  (c) For $k_{0}z_{c}=0$ and $k_{0}a=\pi/4$ the local field quadratures are again maximally correlated, but the Hamiltonian $\mathcal{H}_{\text{scatt}}$ scrambles the array into the dimerized state of Eq.~(\ref{eq:dimer_tot}). In all cases $N_{\text{at}}=6$, $N_{\text{ph}}=0.88$, and $\varphi =0$. }\label{Fig:spatial_correlations}
\end{figure*}

We now turn our attention to the transient dynamics of the array. These dynamics are ruled by the stochastic nature of the field, which determines the polarization decay rates. The connection between array dynamics and steady state---which mirrors the connection between previous studies~\cite{Gardiner_1986} and this work---is made in Figs.~\ref{Fig:time_scales} and~\ref{Fig:spatial_correlations}.

Figure~\ref{Fig:time_scales} shows the evolution towards dimer and melted dimer states for increasing array sizes. In both cases the population $\langle S_{z} \rangle$ is shown to quickly reach its steady state as the atoms dimerize, while the purity takes a longer time that increases with atom number. The stabilization into the dark state follows two different timescales. A fast scale where dissipation brings the array into several dimerized states and a slow scale where it reorganizes into a particular dimer through the Hamiltonian evolution and two non-commuting jump operators. For the melted state, where the system explores a reduced space, the array is quickly brought to the steady state. By comparison, for the dimerized state the array takes a long time to self-organize, a time that increases exponentially with array size. This long timescale is to be compared with the non-radiative and dephasing decays that emerge in an experimental setting. Recent experiments using superconducting qubits work with non-radiative and dephasing decay rates of $\gamma_{\text{nr}}/2\pi\sim 15$kHz and $\kappa_{\phi}/2\pi \sim 100$kHz ($T_{1} = 1.71 \mu s$ and $T_{2} = 0.58 \mu s$)~\cite{Kirchmair_2022}. Considering a typical decay rate of a single-qubit into the waveguide $\gamma/2\pi\sim 15$MHz, the steady states of Fig.~\ref{Fig:time_scales} are reached in a timescale much faster than the decoherence times.

Figure~\ref{Fig:spatial_correlations} shows how the amplification and deamplification of field quadratures cause atomic polarizations to decay at different rates. We plot the evolution and steady state for an array of six atoms whose distances and centers are displaced, thus changing the probed fluctuations from thermal to squeezed. In all cases, each atom is prepared in the state $\vert \psi_{n} \rangle = (\vert g_{n} \rangle + e^{i \pi/4}\vert e_{n} \rangle)/\sqrt{2}$ where the polarizations $\langle S_{x,y} \rangle$ take the same expected value. 

In Figure~\ref{Fig:spatial_correlations}(a) the array is centered on $k_{0}z_{c}=\pi/4$ where left- and right-propagating operators $J_{\phi,s=\pm}$ cancel one another. The array is effectively submerged in an uncorrelated thermal environment. Polarizations decay at the same rate and the population within the $n_{e}$ excited state, $P^{(n_{e})}$, resembles the photon number occupation of a thermal field~\cite{Bergeal_2012}
\begin{equation}\label{eq:thermal}
P^{(n_{e})}_{\text{thermal}} = \frac{x^{n_{e}}}{\sum_{i=1}^{N_{\text{at}}} x^{i}} \, ,
\end{equation}
with $x=N_{\text{ph}}/(N_{\text{ph}}+1)$. The result is to be compared with Fig.~\ref{Fig:spatial_correlations}(b) where the array is centered at $k_{0}z_{c}=0$ and the fields are maximally correlated. As a consequence, the atomic polarizations decay at dramatically different rates and the system settles into a pure state where only even number of excitations are populated [see Eq.~(\ref{eq:collective_agarwal})]
\begin{equation}\label{eq:pop_squeezed1}
P^{(n_{e})}_{\text{squeezed}} = \frac{\vert Y_{\ell,m_{n_{e}}}(\tfrac{\pi}{2},0)\vert^{2} x^{m_{n_{e}}/2}}{\sum_{m_{i}=-\ell}^{\ell} \vert Y_{\ell,m_{i}}(\tfrac{\pi}{2},0)\vert^{2} x^{m_{i}/2} } \, ,
\end{equation}
with $\ell = \tfrac12 N_{\text{at}}$ and $m_{n_{e}}=n_{e}-\tfrac12 N_{\text{at}}$. In both cases the lattice constant is $k_{0}a=2\pi$.

Finally, in Fig.~\ref{Fig:spatial_correlations}(c) the array is centered at $k_{0}z_{c}= \pi /4$ with lattice constant $k_{0}a= \pi /4$. At these points the atoms dimerize, with central and exterior dimers probing two different quadratures. The population is zero for odd excitation numbers, and
\begin{equation}\label{eq:pop_squeezed2}
P^{(n_{e})}_{\text{dimer}} = \binom{\tfrac12 N_{\text{at}}}{ n_{e}} \left(\frac{N_{\text{ph}}+1}{2N_{\text{ph}}+1}\right)^{\tfrac12 N_{\text{at}}} x^{n_{e}} \, 
\end{equation}
for even excitation numbers.

The transition between dimer, melted dimer, and uncorrelated states follows from the correlations of the underlying field. Atomic arrays can act as sensitive probes for quantum fields extending in space, mapping the spatial distribution of the field through decay paths and steady states.
 
\section{Conclusion}\label{Sec:outlook}

Motivated by recent developments in atomic control and environment engineering, we have described the collective radiation of an emitter array into a squeezed vacuum. We found that the correlated fluctuations of the environment drive the array into pure, highly-entangled dark states. These states can be manipulated to display pairwise correlations or all-to-all correlations by changing the relative positions of the emitters. 

The stabilization follows an efficient transfer of correlations from the squeezed field into the array. This was made possible by tailoring the spatial profile of the squeezed modes and array pattern. The former sets a fluctuating background characterized by two-point correlations, while the latter determines how the atoms probe this background field collectively. Following the cascaded system perspective presented here it is possible to move past this white noise limit and study quantum light characterized by correlations to all orders~\cite{Parkins_1988,Noh_2008,Carmichael_2010,Joan_2022}. Such extension sets a promising path to study the transfer of information between light and extended atomic ensembles capable of emulating the correlated photons that form the field. It also raises the possibility to extend the transfer of correlations to atoms coupled at different points of the waveguide~\cite{Kockum_2018,Wang_2021}. For these giant atoms, the concept of subsequent pairs is lost, thus raising the possibility to study degenerate or frustrated states which is a topic of our future research.

Our study is grounded on a simplified picture of superconducting quantum circuits. The extraordinary developments on squeezed environments~\cite{Siddiqi_2013,Siddiqi_2016} together with the exploration of arrays of artificial atoms~\cite{Kirchmair_2022,Orell_2022} make this platform ideal to study the transfer of correlations. The drastic change in the steady state of the atoms, in turn, can be used to probe the extent and decay of spatial noise correlations in these circuits. By probing the collective steady state of two or more atoms it is possible to map how environment correlations propagate. Proper characterization of this noise is central to optimize device performance and can impact circuit design~\cite{Beaudoin_2018,Viola_2020}.

\begin{acknowledgements}
R.~G-J. acknowledges helpful discussions with Z.~K.~Minev and J.~T.~Lee. A.A.-G. and R.G-J. gratefully acknowledge support from the Air Force Office of Scientific Research through their Young Investigator Prize (grant No.~21RT0751). A.A.-G. acknowledges further support from the National Science Foundation through their CAREER Award (No. 2047380), the A.~P.~Sloan foundation, and the David and Lucile Packard foundation. GSA is grateful for support for this work from AFOSR award No. FA9550-20-1-0366.
\end{acknowledgements}



\begin{thebibliography}{1}
\bibitem{us_2023_App} We consider the paths described in Sec.~\ref{Sec:dimerized_chain_one} and evaluate a vanishing escape probability amplitude $c_{esc}=0$ in two cases: (i) $z_{a1},z_{a2}<z_{b1},z_{b2}$ and (ii) $z_{a1}<z_{b1}<z_{a2},z_{b2}$ to give
\begin{align}\label{eq:escape_Ref}
&\sin k_{0}(\bar{z}_{b}-\bar{z}_{a})\cos \tfrac12 k_{0}(z_{rb}+z_{ra})[1 \pm e^{2ik_{0}(\bar{z}_{b}-\bar{z}_{a})}] = \, \nonumber \\
-&\cos k_{0}(\bar{z}_{b}-\bar{z}_{a})\sin \tfrac12 k_{0}(z_{rb}+z_{ra})[1 \mp e^{2ik_{0}(\bar{z}_{b}-\bar{z}_{a})}]  \, .
\end{align}
Upper (lower) signs refers to case (i) [(ii)] and the atomic positions of the $\ell$th pair are written using center and relative coordinates
\begin{align}
& z_{\ell 1} = \bar{z}_{\ell} - \tfrac12 z_{\ell r} \, ,& z_{\ell 2} = \bar{z}_{\ell} + \tfrac12 z_{\ell r} \, . \nonumber
\end{align}
Since all pairs satisfy $\cos 2k_{0}\bar{z}=\pm 1$, Eq.~(\ref{eq:escape_Ref}) excludes nested pairs $(z_{a1} < z_{b1} < z_{b2} < z_{a2})$ with same center $\bar{z}_{b}=\bar{z}_{a}$ for separations $k_{0}a < \pi/2$. The center of at least one pair has to be shifted from the rest. 

Having a shift between centers, condition $\cos 2k_{0} \bar{z}_{\ell} = \pm 1$ can be understood as that of finding the fundamental wavevector $2k_{0}$ connecting all centers. For case (i), where all centers are separated by $2a$ the fundamental mode is $2k_{0} = \pi/2a$ or $k_{0}a = \pi/4$. In all other cases the separations range from $a$ to $a(N_{at}-3/2)$. These separations are commensurate with $a/2$ to give the fundamental mode $k_{0}a = \pi$. Commensurability then requires $k_{0}a= n \pi /4$.

We focus then on possible dimers for $k_{0}a= n \pi/4$. For $k_{0}a = \pi/2$ the interaction matrix $\mathcal{H}_{nm}=\langle e_{n} \vert \mathcal{H}_{\text{scatt}} \vert g_{m} \rangle$ is zero along the diagonal and cycles between $\lbrace 0,\pm \tfrac12 \hbar \gamma \rbrace$ as we advance along rows or columns. Being a symmetric matrix, this cycle is interrupted at the diagonal where it reverses. This change leads to more positive than negative entries, so the only way to reach $c_{esc}=0$ is to cancel the entries above and below the diagonal by pairing neighbouring atoms starting from the boundaries. The same argument follows for $k_{0}a=\pi/4$ after accounting for the sign change between neighboring centers.
\bibitem{us_2022_f} The number of different ways atoms can pair to form a state of excitation $n_{e}$ is
\begin{equation}
{N}_{n_{e}} = \frac{N_{\text{dim}} N_{\text{exc}}(n_{e},n_{g})}{N_{\text{red}}(n_{e},n_{g})}
\end{equation}
where $N_{\text{dim}}$ is the number of possible dimers
\begin{align}\label{eq:extra_dimer}
&N_{\text{dim}} = \frac{1}{(\tfrac12 N_{\text{at}})!}\prod_{n=1}^{\tfrac12 N_{\text{at}}}\binom{2n}{2} = \frac{(N_{\text{at}}!)}{2^{N_{\text{at}}/2} (\tfrac12 N_{\text{at}})!} \, ,
\end{align}
$N_{\text{red}}$ the number of pairs repeating in different dimers 
\begin{align}
&N_{\text{red}}(n_{e},n_{g}) = \frac{1}{2^{(n_{g}+n_{e})/2}} \left( \frac{n_{g}! n_{e}!}{(\tfrac12 n_{g})!(\tfrac12 n_{e})!} \right) \, ,
\end{align}
and $N_{\text{exc}}$ the number of states with the same excitation within each dimer
\begin{equation}
N_{\text{exc}}(n_{e},n_{g}) = \binom{\tfrac12 N_{\text{at}}}{\tfrac12 n_{e}} \, .
\end{equation}
The probability amplitude $c_{e}$ is proportional to the square root of ${N}_{n_{e}}$ after accounting for terms independent of $n_{e}$.
\end{thebibliography}
\end{document}